\documentclass[11pt,twocolumn]{IEEEtrantuhh}
\input epsf
\usepackage{times,mathptm,amsmath}
\usepackage{graphicx}
\usepackage[hang]{subfigure}
\usepackage{epsfig}
\usepackage{enumerate}

\usepackage{setspace}

\def\BibTeX{{\rm B\kern-.05em{\sc i\kern-.025em b}\kern-.08em
    T\kern-.1667em\lower.7ex\hbox{E}\kern-.125emX}}

\begin{document}

\title{Fair Scheduling in OFDMA-based Wireless Systems with QoS Constraints}

\author{
\authorblockN{Tolga Girici\authorrefmark{2}, Chenxi Zhu\authorrefmark{1}, Jonathan R. Agre\authorrefmark{1} and Anthony
Ephremides\authorrefmark{2}}\\
\authorblockA{\authorrefmark{2}
Institute for Systems Research\\
University of Maryland}\\
\authorblockA{\authorrefmark{1}
Fujitsu Labs of America\\
College Park, Maryland 20740}\\
tgirici@glue.umd.edu }

\maketitle
\begin{abstract}
In this work we consider the problem of downlink resource allocation
for proportional fairness of long term received rates of data users
and quality of service for real time sessions in an OFDMA-based
wireless system. The base station allocates available power and
bandwidth to individual users based on long term average received
rates, QoS based rate constraints and channel conditions.  We solve
the underlying constrained optimization problem and propose an
algorithm that achieves the optimal allocation. Numerical evaluation
results show that the proposed algorithm provides better QoS to
voice and video sessions while providing more and fair rates to data
users in comparison with existing schemes.

\end{abstract} \normalsize

\section{Introduction}

Recent  IEEE 802.16 standard defines the air interface and medium
access control (MAC) specifications for wireless metropolitan area
networks.  Such networks intend to provide high speed voice, data
and on demand video streaming services for end users. IEEE 802.16
standard is often referred to as WiMax and it provides substantially
higher rates than cellular networks. Besides it eliminates the
costly infrastructure to deploy cables, therefore it is becoming an
alternative to cabled networks, such as fiber optic and DSL systems
\cite{Ekl02},\cite{Gho05}.

Transmissions in IEEE 802.16 networks is based on Orthogonal
frequency division multiplexing (OFDM), which is a multicarrier
transmission technique that is proposed for high speed wireless
transmission.  It is based on a large number of orthogonal
subchannels, each working at a different frequency. Within OFDMA
framework, the resource allocated to the users come in three
dimensions: time, frequency and power. This requires the scheduler
to operate with higher degree of freedom but also makes the notion
of resource fairness  obsolete and makes the problem more involved.

There are  three main issues that need to be considered in multiple
access resource allocation, which are spectral efficiency, fairness
and Quality of Service (QoS). These are often contradicting aims and
achieving these in an OFDMA based system is a challenging task.
OFDMA-based scheduling for systems with heterogeneous traffic has
not been studied sufficiently. Recently in \cite{Kim05},
\cite{Son05}, \cite{Zhu06}, proportional fair scheduling was studied
for multicarrier systems. However, the scheduling rules in these
works do not apply sufficiently to different QoS requirements and
heterogeneous traffic. Besides they only provide fairness for short
term received rate, instead of long term. In this work our goal is
to find algorithms that provide proportional fairness to data users
and QoS for real time sessions such as voice and video streaming.

\section{System Model}

We consider a downlink system, where a base station transmits to N
users. Each user is assumed to demand a single type of traffic. Let
$W$ and $P$ denote the total bandwidth and power, respectively.
Total bandwidth $W$ is divided into $N_{sub}$ subchannels of length
$W_{sub}$, each consisting of a group of carriers.  We consider PUSC
as the  mode of subchannelization, which is more suitable for mobile
users than AMC.  It provides frequency diversity and inter-cell
interference averaging. This minimizes the performance degradation
due to fast fading characteristics of mobile environments. We assume
flat fading, which is a reasonable assumption with PUSC.

The noise and interference power density is $N_0$, and the channel
gain averaged over the entire band from the BS to user $i$ at time
$t$ is $h_i(t)$, where $h_i(t)$ includes path loss, shadowing
(lognormal fading) and fast fading. There are three classes of
users. Users in the classes $U_D$, $U_S$ and $U_V$ demand data,
voice and video traffic, respectively. The system that we consider
is time slotted with frame length $T_s$. The scheduler makes a
resource allocation decision at each frame. Active period in a voice
conversation, streaming duration and file size are both very long
with respect to the frame size. Therefore it is realistic that
during the course of optimization the number of active sessions are
fixed.


In problem formulation we will adopt the Shannon channel capacity
for AWGN channel as a function for bandwidth and transmission power
assigned to user $i$:
\begin{equation}
r_i(w_i(t),p_i(t))=w_i(t)\log\left(1+\beta\frac{p_i(t)h_i(t)}{N_0w_i(t)}\right)\label{capacity}
\end{equation}
The reason for using Shannon capacity is its simplicity, and it also
approximates the above set of rate-SINR relation with $\beta=0.25$.
The parameter $0<\beta<1$ compensates the rate gap between Shannon
capacity and rate achieved by practical modulation and coding pairs
listed in IEEE 802.16a/e standards. After finding power and
bandwidth pairs by solving the convex programming problem, we
quantize the SINRs to the thresholds determined in the standard. In
this paper, we propose a Fair and QoS-based Power and Subchannel
Allocation (FQPSA) Algorithm. Objective of the algorithm is to
maximize proportional fairness for data users subject to rate
constraints of real time users selected each frame according to a
metric.

\section{Problem Formulation}

\subsection{Data Traffic}
The objective for the data users is to optimize log sum of the
exponential averaged rates of the users  Average received user rates
are updated at each frame according to the exponential averaging
formula:
\begin{equation}
R_i(t)=\alpha_i R_i(t-1)+ (1-\alpha_i)r_i(w_i(t),p_i(t)), \forall
i,t \label{update}
\end{equation}
This way we consider both current rate as well as rates given to the
user in the past. Observed at time $t$, the highest consideration is
given to the current rate $r(t)$, and the rates received at the past
$t-1$, $t-2$,... carry diminishing importance.
\begin{multline}
C(\mathbf{R}(t))=\sum_{i=1}^{N}\log R_i(t)=\\
 C(\mathbf{R}(t-1))+\sum_{i=1}^{N}\log\left(\alpha +
\frac{\overline{\alpha}r_i(w_i(t),p_i(t))}{R_i(t-1)}\right)\label{logsum}
\end{multline}

As a matter of fact, we limit ourself to \textit{greedy} schemes in
the sense that at frame $t$, we try to maximize the proportional
fair capacity $C(\mathbf{R}(t))$ without considering the future
frames $t+1$,$t+2$, etc. Only the second term in (\ref{logsum})
needs consideration. Here  the log-sum can be replaced with product.


\subsection{Real Time Traffic}

The Base Station first chooses a number of voice and video users
according to a metric that reflects current channel condition, head
of line packet delay and long term received rate. Then for the
selected users the number of bits to be transmitted in the current
frame is determined. These procedures are explained in \cite{Gir06},
in detail. Let $U_R'$ be the set of real time sessions selected and
let $r_i^c$ be the rate requirement of real time session $i$.


The resulting optimization problem considering all types of sessions
is as follows: Find, \footnote{Here $p_i$,$w_i$, $n_i$ are the
values at time $t$. The time index is not shown for convenience.}:
\begin{equation}
(\mathbf{p^*,w^*})\nonumber =\arg\max_{\mathbf{p},\mathbf{w}}
\prod_{i\in
  U_D}{\left(\alpha_i+\frac{\overline{\alpha_i}w_i\log\left({1+\frac{ p_i}{n_iw_i}}\right)}{R_i}\right)}\label{const_prob}
\end{equation}
subject to,
\begin{eqnarray} \sum_{i\in U_D\cup U_R'}{p^*_i}&\leq& P \label{power_const}\\
\sum_{i\in U_D\cup U_R'}{w^*_i}&\leq& W \label{band_const}\\
w^*_i\log\left({1+\frac{ p^*_i}{n_i w^*_i}}\right)&\geq & r_i^c, i\in U_R'\label{rate_const}\\
p^*_i, w^*_i &\geq& 0, \forall i\in U_D\cup U_R'\label{gr_than_0}
\end{eqnarray} where $n_i=\frac{N_0}{\beta h_i}$.

\section{Solution of the Optimization Problem}

The objective function (\ref{const_prob}) is an increasing function
of $\mathbf(w,p)$, therefore the maximum is achieved only when the
constraints (\ref{power_const}, \ref{band_const}, \ref{rate_const})
are all met with equality. For this reason we can replace these
inequalities with equalities in the discussion below.

Actually there is no guarantee that a solution can be found that
satisfies both rate and power constraints, given the rate
requirements and channel conditions. The rate requirement can be too
high that it may be impossible to satisfy with the given channel
conditions. To start with, we assume that the problem is
\textit{feasible}. We will discuss about how to detect infeasibility
of the problem and what to do in that case in the next section. We
can write the Lagrangian of the problem as:



\begin{multline}
L(\mathbf{w},\mathbf{p},\lambda_p,\lambda_w,\mathbf{\lambda}^r)=
 \prod_{i\in  U_D}
{\left(\alpha_i+\frac{\overline{\alpha_i}w_i\log\left({1+\frac{
p_i}{n_iw_i}}\right)}{R_i}\right)}\\+ \lambda_p\left(P-\sum_{i\in
U_D\cup U_R'}p_i\right)+\lambda_w\left(W-\sum_{i\in U_D\cup
U_R'}w_i\right)
\\+\sum_{i\in U_R'}\lambda_i^r\left(w_i\log\left(1+\frac{
p_i}{n_iw_i}\right)-r_i^0\right).\label{lagrangian}
\end{multline}
From now on we assume that all rates are in nats/sec and logarithms
are natural. Taking the derivatives of
$L(\mathbf{p},\mathbf{w},\lambda_p, \lambda_w, \mathbf{\lambda}^r
)$,
 we get the following:

\subsection{ For users $i\in U_D$}
\begin{multline}
\left.\frac{\partial L(\mathbf{p},\mathbf{w},\lambda_p, \lambda_w,
\mathbf{\lambda}^r )}{\partial
p_i}\right|_{(\mathbf{p}^*,\mathbf{w}^*)} = 0~~ \Rightarrow\\
\frac{L^*}{\lambda_p}= (n_iw_i^*+ p_i^*
)\left(\frac{R_i\widetilde{\alpha}_i}{w_i^*}+ \log\left(1+\frac{
p_i^*}{n_iw_i^*}\right)\right) \label{data_eq1}\end{multline}

\begin{multline}
\left.\frac{\partial L(\mathbf{p},\mathbf{w},\lambda_p, \lambda_w,
\mathbf{\lambda}^r )}{\partial
w_i}\right|_{(\mathbf{p}^*,\mathbf{w}^*)}=0 ~~\Rightarrow\\
\frac{L^*}{\lambda_w}= \frac{(n_iw_i^*+
p_i^*)\left(R_i\widetilde{\alpha}_i+w_i^*\log\left(1+\frac{
p_i^*}{n_iw_i^*}\right) \right)}{(n_iw_i^*+ p_i^*)\log\left(1+\frac{
p_i^*}{n_iw_i^*}\right)-p_i}\label{data_eq2}
\end{multline}where
$\widetilde{\alpha}_i=\frac{\alpha_i}{1-\alpha_i}$. Let
$\Lambda_p=\frac{L^*}{\lambda_p}$ and
$\Lambda_w=\frac{L^*}{\lambda_w}$. By dividing (\ref{data_eq1}) with
(\ref{data_eq2}) we can write for all $i\in U_D $:

\begin{equation}
\frac{\Lambda_p}{\Lambda_w}=\Lambda_a=n_i\left(\left(1+x_i^*\right)\log\left(1+x_i^*\right)-x_i^*\right),
\label{data_eq3}
\end{equation}where $x_i^*=\frac{ p_i^*}{n_iw_i^*}$ denotes the
optimal \textit{effective} SINR, which is the SINR multiplied by the
SINR gap parameter $\beta$.

\subsection{ For users $i\in U_R'$}

\begin{equation}
\left.\frac{\partial L(\mathbf{p},\mathbf{w},\lambda_p, \lambda_w,
\mathbf{\lambda}^r )}{\partial
p_i}\right|_{(\mathbf{p}^*,\mathbf{w}^*)} \Rightarrow
\frac{\lambda_p}{\lambda_i^v} =\frac{1}{n_i}\frac{1}{1+\frac{
p_i^*}{n_iw_i^*}} \label{voice_eq1}
\end{equation}
\begin{multline}
\left.\frac{\partial L(\mathbf{p},\mathbf{w},\lambda_p, \lambda_w,
\mathbf{\lambda}^r )}{\partial
w_i}\right|_{(\mathbf{p}^*,\mathbf{w}^*)} \Rightarrow\\
\frac{\lambda_w}{\lambda_i^r} = \log\left(1+\frac{
p_i^*}{n_iw_i^*}\right)-\frac{\frac{ p_i^*}{n_iw_i^*}}{1+\frac{
p_i^*}{n_iw_i^*}}\label{voice_eq2}
\end{multline}

With $\frac{\Lambda_p}{\Lambda_w}=\frac{\lambda_w}{\lambda_p}$,
dividing equation (\ref{voice_eq2}) to (\ref{voice_eq1}) for all
$i\in U_R'$ again gives (\ref{data_eq3}).

By dividing (\ref{voice_eq2}) to (\ref{voice_eq1}) we can eliminate
$\lambda_i^v$'s from the problem. It is worth noting that we get the
same relation between $\Lambda_a/n_i$ and $x_i$ for both data and
real time sessions. At this point it is convenient to define the
function $f_a(x)$ as:
\begin{equation}
f_a(x)=(1+x)\log(1+x)-x.
\end{equation}
Then we have,
\begin{equation}
x_i=f_a^{-1}(\Lambda_a/n_i), \forall i \in U_D\cup
U_R'\label{lambda_a_sinr}
\end{equation} Signal to noise ratio ($x_i$) is a monotonic increasing function of
$\Lambda_a$ for all users $i\in U_D\cup U_R'$.


For real time users we also have:
\begin{multline}\left.\frac{\partial L(\mathbf{p},\mathbf{w},\lambda_p,
\lambda_w, \mathbf{\lambda}^r )}{\partial
\lambda_i^v}\right|_{(\mathbf{p}^*,\mathbf{w}^*)}=0 ~~ \Rightarrow\\
r_i^0 =w_i^*\log\left(1+\frac{p_i^*}{n_iw_i^*}\right), \forall i\in
U_R'\label{voice_eq3}\end{multline}

\subsection{ For all nodes $i\in U_D\cup U_R'$}
\begin{equation}
\left.\frac{\partial L(\mathbf{p},\mathbf{w},\lambda_p, \lambda_w,
\mathbf{\lambda}^r )}{\partial
\lambda_p}\right|_{(\mathbf{p}^*,\mathbf{w}^*)}=0 ~~ \Rightarrow P =
\sum_{i\in U}p_i^* \label{all_eq1}\end{equation}
\begin{equation}
\left.\frac{\partial L(\mathbf{p},\mathbf{w},\lambda_p, \lambda_w,
\mathbf{\lambda}^r )}{\partial
\lambda_w}\right|_{(\mathbf{p}^*,\mathbf{w}^*)}=0 ~~ \Rightarrow W =
\sum_{i\in U}w_i^*\label{all_eq2}
\end{equation}
From Equation (\ref{data_eq1}) for data users we can write:
\begin{eqnarray}
\frac{[\Lambda_p-n_i(1+x_i^*)R_i\widetilde{\alpha}_i]^+}{\log(1+x_i^*)(1+x_i^*)n_i}&=&w_i^*,
i\in U_D \label{lambdap1}\\
\frac{[\Lambda_p-n_i(1+x_i^*)R_i\widetilde{\alpha}_i]^+x_i^*}{\log(1+x_i^*)(1+x_i^*)}&=&p_i^*
, i\in U_D\label{lambdap3}
\end{eqnarray}
The $[.]^+$ operator in Equations (\ref{lambdap1}),(\ref{lambdap3})
guarantees that $w_i,p_i\geq 0$ for all users. Given $\Lambda_p$ and
$\Lambda_a$ we can compute the power and bandwidth for users $i\in
U_D$ using (\ref{lambdap1}) and (\ref{lambdap3}). Given $\Lambda_a$,
we can calculate the power and bandwidth for users $i\in U_R'$ using
(\ref{voice_eq3}). Please note that just from (\ref{voice_eq3}),
(\ref{lambdap1}) and (\ref{lambdap3}), the bandwidth and power
constraints (\ref{band_const}) (\ref{power_const}) are \textit{not}
necessarily satisfied. We need to find the right $\Lambda_a$ and
$\Lambda_p$ so that the power and bandwidth constraints are
satisfied.  Let $S_p(\Lambda_a,\Lambda_p)$ and
$S_w(\Lambda_a,\Lambda_p)$ be the total bandwidth and total power
corresponding to $\Lambda_a$ and $\Lambda_p$:
\begin{multline}
S_w(\Lambda_a,\Lambda_p)=\sum_{i\in U_D\cup
U_R'}{w_i(\Lambda_a,\Lambda_p)}=\\\sum_{i\in
U_D}\frac{[\Lambda_p-n_i(1+x_i)R_i\widetilde{\alpha}_i]^+}{\log(1+x_i)(1+x_i)n_i}+\sum_{i\in
U_R'}\frac{r_i^c}{\log{(1+x_i)}}\label{lambdap2}\end{multline}
\begin{multline}
S_p(\Lambda_a,\Lambda_p)=\sum_{i\in U_D\cup
U_R'}{p_i(\Lambda_a,\Lambda_p)}=\\\sum_{i\in
U_D}\frac{[\Lambda_p-n_i(1+x_i)R_i\widetilde{\alpha}_i]^+x_i}{\log(1+x_i)(1+x_i)}+\sum_{i\in
U_R'}\frac{r_i^cx_in_i}{\log{(1+x_i)}}\label{lambdap4}
\end{multline}

where $x_i=f_a^{-1}(\Lambda_a/n_i)$ is the SNR of user $i$. As a
result, the problem is finding $\Lambda^*_a$ and $\Lambda^*_p$ such
that
\begin{eqnarray}
S_w(\Lambda^*_a,\Lambda^*_p)&=&W\label{sum_w}\\
S_p(\Lambda^*_a,\Lambda^*_p)&=&P\label{sum_p}
\end{eqnarray}using Equations (\ref{lambda_a_sinr}),(\ref{sum_w})
and (\ref{sum_p}).  Note that although $\Lambda_a$ and $\Lambda_p$
are independent variables that determine power and bandwidth for
each node, they become dependent when the power and bandwidth
constraints (\ref{sum_w}) (\ref{sum_p}) need to be satisfied.

\subsection{Feasibility of the Solution}

From (\ref{lambdap1}) and (\ref{lambdap3}), $\Lambda_p=0$
corresponds to the case that no bandwidth and power is allocated to
data sessions. If the problem is feasible, then there exists a
$\Lambda_a$ that satisfies the following:
\begin{eqnarray}
S_w(\Lambda_a,0)&=&W,\label{feasible1}\\
S_p(\Lambda_a,0)&\leq&P.\label{feasible2}
\end{eqnarray}

Let $\Lambda_a^0$ be the smallest $\Lambda_a$ that satisfies
(\ref{feasible1}) and (\ref{feasible2}). In the report \cite{Gir06}
we proved the existence and uniqueness of $\Lambda_a^0$ for a
feasible problem using monotonicity properties of the functions
$S_w(\Lambda_a,\Lambda_p)$ and $S_p(\Lambda_a,\Lambda_p)$.


\section{Proposed Algorithm}

Using (\ref{lambda_a_sinr}),(\ref{lambdap2}) and (\ref{lambdap4}) we
can develop an algorithm to determine the power and bandwidth
allocation. The algorithm is also able to detect infeasibility if
there no solution exists.

\underline{\textbf{Algorithm:}}
\begin{enumerate}

\item Find $\Lambda_a^0$. If $S_p(\Lambda_a^0,0)<P$ then the problem is feasible. If not, find
the real time session that consumes the largest power and decrease
its rate. Repeat Step 1 and 2 until feasibility is achieved.

\item Using two nested binary searches find the unique $(\Lambda_a^*,
\Lambda_p^w(\Lambda_a^*))$ that satisfies power and bandwidth
constraints.

\item Compute optimal power and bandwidth using the optimal $(\Lambda_a^*,
\Lambda_p^w(\Lambda_a^*))$.
\end{enumerate}

Let us summarize the nested binary search. We start from a
$\Lambda_a$, then we find the $\Lambda_p^w(\Lambda_a)$ that
satisfies the bandwidth constraint. If the power constraint is
exceeded then we decrease $\Lambda_a'$, else increase it, and search
for $\Lambda_p^w(\Lambda_a')$. We continue until both constraints
are satisfied.

Since the problem is convex (proved in \cite{Gir06}), the problem
has a unique solution.  We need the monotonicity property of
function $S_p(\Lambda_a,\Lambda_p^w(\Lambda_a))$ (sum of powers) as
a function of $\Lambda_a$ in order to be able to find it using
binary search. Those properties are also verified in \cite{Gir06}.

\section{Performance Evaluation}

For the numerical evaluations we equally divide the users to 5
classes according to the distances, 0.3,0.6,0.9,1.2,1.5 km. We use
the parameters in Table \ref{simulation_parameters}.
\begin{table}[htb]\centering
\footnotesize
\begin{tabular}{|c|c|}
\hline
 \textbf{Parameter}&\textbf{Value}\\\hline\hline
Cell radius & 1.5km\\\hline

User Distances & 0.3,0.6,0.9,1.2,1.5 km \\\hline

Total  power (P)&20 W \\\hline

Total bandwidth (W) &8.3 MHz\\\hline

Frame Length & 1 msec \\\hline

Voice Traffic & CBR 32kbps \\\hline

Video Traffic &  802.16 - 128kbps \\\hline

FTP File &  5 MB \\\hline

AWGN p.s.d.($N_0$)&-174dBm/Hz  \\\hline 

Pathloss exponent ($\gamma$)& 3.5\\\hline

$\psi_{DB}\sim N(\mu_{\psi_{dB}},\sigma_{\psi_{dB}})$ & N(0dB,8dB)
\\\hline

 Coherent Time (Fast/Slow Fading) & (5msec/300msec.)  \\\hline

Pathloss(dB, d in meters)& $-31.5- 35\log_{10}d+\psi_{dB}$\\\hline
\end{tabular}
\caption{Simulation Parameters}\label{simulation_parameters}
\end{table}

We will use M-LWDF-PF scheme as benchmark in our simulations. In
this scheme at each frame the user maximizing the following quantity
transmits.
\begin{equation}a_i D_i^{HOL}(t)\log(1+\beta P/N_0W),\label{lwdf}\end{equation}where
$D_i^{HOL}(t)$ is the head of line packet delay and $r_i(t)$ is the
channel capacity of user i at frame t. The parameter $a_i$ is a
positive constant. The constant $a_i$ is defined as
$a_i=-\frac{\log(\delta_i)}{D_i^{max}R_i(t)}$ , which is referred to
as M-LWDF-PF \cite{And01} \cite{Son05}. M-LWDF-PF can be adapted to
OFDMA systems as follows. Power is distributed equally to all
subchannels. Starting from the first subchannel , the subchannel is
allocated to the user maximizing (\ref{lwdf}). Then the received
rate $R(t)$ is updated according to (\ref{update}). All the
subchannels are allocated one-by-one according to this rule.  Delay
exceeding probability is taken as $\delta_i=0.05$ for all users.
Delay constraint for voice, video and data users are 0.1, 0.4 and 1
seconds, respectively.


Tables \ref{increasing_voice}, \ref{increasing_video} and
\ref{increasing_data} show the effects of increasing the number of
voice, video streaming and data users, respectively. In these tables
we observe that FPQS Algorithm is better than M-LWDF algorithm in
terms of both delay performance and data performance.  With FQPS
delay for voice and video sessions stay within acceptable bounds,
while with M-LWDF, it exceeds the bounds for the user at the cell
edge when $V\geq 30$ or $S>40$. Besides, FQPSA provides at least 10
percent increase in total throughput. Total throughput decreases
linearly with increasing number of voice and video users. Although
10 voice users adds up to 0.32 Mbps, adding 10 users decreases the
total throughput approximately by 1.2-1.4 Mbps. This is because
voice has a very strict delay requirement and a voice session may
have to be transmitted despite bad channel conditions. Throughput
for LWDF decreases with a little bit slower rate but that reflects
to the voice and video performance negatively. Log-sum performance
of FQPSA is also better than that of M-LWDF, which shows that our
algorithm provides fairness.
\begin{table}[htb]\centering
\footnotesize
\begin{tabular}{|c|c|c|c|c|c|c|}
\hline

V = & 10&20&30&40&50&60\\\hline\hline

FQPS Voice (G)&36 &33&30&32&31&31\\\hline

FQPS Voice (B)&88 &66&62&81&82&80\\\hline

LWDF Voice (G)&41& 41& 41& 41& 41& 41\\\hline

LWDF Voice (B)&89& 91& 93& 110& 133& 151\\\hline

FQPS Video (G)&148 &132&130&132&132&131\\\hline

FQPS Video (B)&295 &271&250&263&266&252\\\hline

LWDF Video (G)& 180& 187& 192& 200& 202& 208\\\hline

LWDF Video (B)&384& 399& 408& 410& 423& 435\\\hline

FQPS (Mbps)&26.9 &25.2&23.8&22.7&21.4&20.1\\\hline

FQPS log-sum&280& 279& 278& 277& 276& 275\\\hline

LWDF (Mbps)&23.9& 22.7& 21.6& 20.5& 19.4& 18.5\\\hline

LWDF Log-sum&275& 278& 276& 274& 272& 269\\\hline

\end{tabular}
\caption{Effects of increasing $\#$ of Voice users for $S=20,D=20$.
Units: Delay (msec), t.put (Mbps)}\label{increasing_voice}
\end{table}
\begin{table}[htb]\centering
\footnotesize
\begin{tabular}{|c|c|c|c|c|c|c|}
\hline
 S = & 10&20&30&40&50&60\\\hline\hline

FQPS Voice (G)&31 &33&39&43&49&51\\\hline

FQPS Voice (B)&57 &66&71&87&93&108\\\hline

LWDF Voice (G)&41& 41& 41& 41& 41& 41\\\hline

LWDF Voice (B)&86& 91& 94& 101& 142& 117\\\hline

FQPS Video (G)&122 &132&150&167&174&190\\\hline

FQPS Video (B)&217 &271&275&301&338&392\\\hline

LWDF Video (G)&171& 187& 190& 194& 202& 223\\\hline

LWDF Video (B)&395& 399& 399& 414& 463& 519\\\hline

FQPS D(Mbps)&27.2 &25.2&23.5&21.5 &19.5&17.5\\\hline

FQPS Log-sum&281& 279& 278& 276& 274& 272\\\hline

LWDF (Mbps)&25.1& 22.7& 20.7& 18.6& 16.7& 14.6\\\hline

LWDF log-sum&279& 277& 277& 276& 275& 273\\\hline

\end{tabular}
\caption{Effects of increasing $\#$ of Video Streaming users for
$D=20,V=20$. Units: Delay (msec), t.put
(Mbps)}\label{increasing_video}
\end{table}


In Table \ref{increasing_data} we can observe the effects of
increasing the number of data users.We observe that delay for both
voice and video streaming sessions stay approximately constant.
Delay performance is much better than that of M-LWDF algorithm. Data
performance is 10 percent better than M-LWDF. Total throughput
increases with number of data users, but the increase diminishes as
D increases.
\begin{table}[htb]\centering
\footnotesize
\begin{tabular}{|c|c|c|c|c|c|c|}
\hline

D = & 20&30&40&50&60&70\\\hline\hline

FQPS Voice (G)&38 &33&34&34&36&39\\\hline

FQPS Voice (B)&72 &66&69&71&71&91\\\hline

LWDF Voice (G)& 41& 47& 50& 57& 61& 67\\\hline

LWDF Voice (B)&91& 106& 126& 146& 170& 182\\\hline

FQPS Video (G)&155 &132&143&145&142&168\\\hline

FQPS Video (B)&278 &271&246&270&270&323\\\hline

LWDF Video (G)&187& 232& 265& 297& 335& 357\\\hline

LWDF Video (B)&399& 460& 537& 557& 612& 656\\\hline

FQPS (Mbps)&24.5 &25.2&25.8&26.1&26.2&26.6\\\hline

FQPS Log-sum&146& 279& 407& 532& 654& 775\\\hline

LWDF (Mbps)&22.7& 23.3& 23.7& 24.0& 24.1& 24.3\\\hline

LWDF Log-sum&146& 278& 405& 530& 653& 772\\\hline
\end{tabular}
\caption{Effects of increasing $\#$ of Data users for $S=20,V=20$.
Units: Delay (msec), t.put (Mbps)}\label{increasing_data}
\end{table}
\bibliographystyle{unsrt}
\bibliography{Multicarrier_Proportional_Conference}
\end{document}